\documentclass[12pt]{article}

\usepackage{amsmath, amssymb, amsfonts}
\usepackage{bm}
\usepackage{graphicx}
\usepackage{hyperref}
\usepackage{geometry}
\usepackage{setspace}
\usepackage{caption}

\renewcommand{\thefootnote}{\dagger}

\title{Relaxation Times for Nonextensive Systems Using Gradient Flow for the Maximization of Tsallis Entropy: An Application to Financial Market Dynamics}

\author{Sandhya Devi\thanks{Shell International Exploration and Production Co. (Retired)}}

\date{}

\begin{document}
\maketitle

\begin{center}
509 6th Ave S, Edmonds, WA 98020, United States of America\\
Email: sdevi@entropicdynamics.com
\end{center}

\renewcommand{\thefootnote}{\arabic{footnote}}

\begin{abstract}
In this work, we develop a method to estimate the relaxation time (the time required to reach equilibrium) of a nonextensive system such as financial market dynamics, using a Euclidean Gradient Flow (EGF) framework for the maximization of Tsallis entropy. The equilibrium state is defined as the maximum-entropy state. Specifically, the dynamics are expressed in terms of the time variations of the $q$-Gaussian parameters---the entropic index $q$ and the inverse temperature beta---under the constraint that the distributions remain $q$-Gaussian at all times. We show that, for nonextensive systems, the relaxation times are longer than those obtained from the maximization of Shannon entropy, indicating that predictions over longer times are possible.
\end{abstract}

\noindent\textbf{Keywords:} Euclidean gradient flow, nonextensive statistics, Tsallis entropy, maximum Tsallis entropy, nonlinear dynamics.

\section{Introduction}

In 1948, Shannon~\cite{Shannon1948} introduced an information-theoretic approach that established the concept of information entropy. In 1983, Jaynes~\cite{Jaynes1983} showed that maximizing Shannon entropy with constraints on the first three moments yields a Gaussian distribution. These works provided mathematical tools that could be applied to financial models~\cite{Fama1965} based on the Efficient Market Hypothesis (EMH)~\cite{Bachelier1964}, which assumes: (a) investors have all available information and independently make rational decisions; (b) markets react quickly to new information, reaching equilibrium rapidly; (c) in this equilibrium state, market behavior follows a random walk; and (d) market returns form high-entropy systems and are therefore not predictable. In such a system, extreme changes are expected to be very rare.

In reality, however, the financial market is a highly nonlinear and complex system driven by the decisions of interacting agents (e.g., herding behavior), traders who speculate, and individuals who react impulsively to limited information. Such collective or chaotic behavior can lead to wild swings in the system, resulting in phenomena such as phase transitions, bubbles, crashes~\cite{Sornette2003}, superdiffusion, and more. Moreover, the equilibrium distributions of market returns are not Gaussian~\cite{Mandelbrot1997} but instead exhibit sharp peaks and fat tails.

Several studies~\cite{Tsallis2003,Cortines2007,Osorio2004,TsallisBook2000,Devi2017,TsallisBukman1996} indicate that these issues can be addressed using statistical methods based on Tsallis entropy (also known as $q$-entropy),  a generalization of Shannon entropy designed for nonextensive systems. These formulations were originally developed to study classical and quantum chaos, physical systems far from equilibrium (such as turbulent flows), and long-range interacting Hamiltonian systems. In recent years, there has been considerable interest in applying these methods to financial market dynamics as well. Such applications fall under the field of econophysics~\cite{MantegnaStanley2000}.

One of the assumptions of the Efficient Market Hypothesis is that the relaxation time---the time required to reach the equilibrium or maximum-entropy state---is very short (assumption (b)). Consequently, prediction of market behavior is considered possible only over very short time intervals. In this paper, we investigate how the relaxation time changes when Tsallis statistics is used. This is carried out within the framework of Euclidean Gradient Flows (EGF)~\cite{Santambrogio2017,Butcher2016} (the term ``Euclidean'' refers to flat-space gradient flows, as opposed to curved-space formulations~\cite{Ambrosio2008}), under the constraint that the distributions remain $q$-Gaussian at all times. The dynamics arise from time variations in the entropic index $q$ and the inverse temperature $\beta$. The EGF equations are obtained from a variational principle that optimizes a free-energy functional (such as entropy) with a penalty term representing the distance between parameter values at successive time steps. The assumption that the distributions remain $q$-Gaussian during the evolution is supported by several studies~\cite{Borland1998,Zanette1999,MichaelJohnson2003}.

The organization of the paper is as follows. Section~2 discusses the EGF formulation for the maximization of both Shannon and Tsallis entropies. Section~3 addresses the numerical aspects of solving the highly nonlinear parameter dynamics and presents the results. A comparison of the relaxation times obtained from Shannon entropy (EMH case) and Tsallis entropy is provided, along with comparisons to empirical data. Section~4 presents the summary and conclusions.

\section{Theory}

\subsection{Euclidean Gradient Flow (EGF)}

In the simple EGF approach \cite{Santambrogio2017, Butcher2016, Ambrosio2008}, 
a free energy functional such as entropy is optimized by a steepest descent in the parameter space. 
If we consider discrete time steps and the free energy functional as the entropy, the parameter dynamics 
for a set of parameters $\{\theta\}$ is obtained by maximizing/minimizing the functional
\begin{equation}
\|\theta - \theta_k\|^2 + \tau F(p)
\tag{1}
\end{equation}
with respect to $\theta$.  
Here $F(p)$ is the free energy functional, $p$ is the probability density function (PDF), 
and $k$ is the $k$th time step. $\tau$ is a constant.

For example, for a two-parameter case $(\theta_1,\theta_2)$, assuming $F(p)$ is to be maximized, we obtain
\begin{equation}
\frac{\partial}{\partial \theta_1}
\left[
(\theta_1 - \theta_{1,k})^2 + (\theta_2 - \theta_{2,k})^2 + \tau F(p)
\right] = 0
\tag{2}
\end{equation}
\begin{equation}
\frac{\partial}{\partial \theta_2}
\left[
(\theta_1 - \theta_{1,k})^2 + (\theta_2 - \theta_{2,k})^2 + \tau F(p)
\right] = 0
\tag{3}
\end{equation}

which gives
\begin{equation}
\theta_{1,k+1} = \theta_{1,k} + \tau \frac{\partial F(p)}{\partial \theta_1},
\tag{4}
\end{equation}
\begin{equation}
\theta_{2,k+1} = \theta_{2,k} + \tau \frac{\partial F(p)}{\partial \theta_2}.
\tag{5}
\end{equation}

Here the factor 2 has been absorbed into $\tau$.  
If we write $\tau = \gamma \Delta t$, where $\gamma$ is a constant and $\Delta t$ is the time discretization, 
then in the limit $\Delta t \to 0$, equations (4) and (5) yield the gradient flow differential equations
\begin{equation}
\frac{\partial \theta_1}{\partial t} = \gamma \frac{\partial F(p)}{\partial \theta_1},
\tag{6}
\end{equation}
\begin{equation}
\frac{\partial \theta_2}{\partial t} = \gamma \frac{\partial F(p)}{\partial \theta_2}.
\tag{7}
\end{equation}

In general, these are nonlinear equations. The parameter $\gamma$ is called the mobility parameter, which must be estimated by fitting to data.

\subsection{Maximization of Shannon Entropy $S_{\mathrm{sh}}$}

In this case,
\begin{equation}
F(p) = \int P(x)\,\log\!\left(\frac{1}{P(x)}\right)\,dx.
\tag{8}
\end{equation}

Assuming a Gaussian distribution
\begin{equation}
P(x) = \sqrt{\frac{\beta_g}{\pi}}\,\exp\!\left[-\beta_g (x - \mu_g)^2\right],
\tag{9}
\end{equation}
with $\mu_g$ as the mean and $\beta_g = 1/(2\sigma^2)$, where $\sigma$ is the standard deviation.

The EGF equations for $\mu_g$ and $\beta_g$ are
\begin{equation}
\frac{\partial \mu_g}{\partial t} = \gamma_g \frac{\partial F(p)}{\partial \mu_g},
\tag{10}
\end{equation}
\begin{equation}
\frac{\partial \beta_g}{\partial t} = \gamma_g \frac{\partial F(p)}{\partial \beta_g}.
\tag{11}
\end{equation}

Using (9), one can show that
\begin{equation}
F(p) = S_{\mathrm{sh}} = -\int P(x)\log P(x)\,dx = \frac{1}{2}\log\!\left(\frac{\pi}{\beta_g}\right).
\tag{12}
\end{equation}

Thus the EGF equations become
\begin{equation}
\frac{\partial \mu_g}{\partial t} = 0,
\tag{13}
\end{equation}
\begin{equation}
\frac{\partial \beta_g}{\partial t} = -\frac{\gamma_g}{2\beta_g}.
\tag{14}
\end{equation}

The solution of (14) is
\begin{equation}
\beta_g(t) = \sqrt{\beta_{0g}^2 - \gamma_g t},
\tag{15}
\end{equation}
where $\beta_{0g} = \beta_g(t=0)$.  
For $\gamma_g > 0$, this is a monotonically decreasing function of $t$.  
Since $\beta_g \ge 0$, equation (15) gives an upper limit for $t$:
\begin{equation}
T_g = \frac{\beta_{0g}^2}{\gamma_g}.
\tag{16}
\end{equation}

At $t = T_g$, the standard deviation becomes infinite (maximum uncertainty), corresponding to the maximum entropy state.

\subsection{Maximization of Tsallis Entropy $S_q$}

The free energy functional is
\begin{equation}
F(p) = S_q = \sum_i P_i \ln_q\!\left(\frac{1}{P_i}\right),
\tag{17}
\end{equation}
where the $q$-logarithm is defined as
\begin{equation}
\ln_q(x) = \frac{x^{1-q} - 1}{1 - q}.
\tag{18}
\end{equation}

Substituting (18) into (17), we obtain
\begin{equation}
S_q = \frac{1 - \sum_i P_i^q}{q - 1}.
\tag{19}
\end{equation}

Unlike Shannon entropy, Tsallis entropy is nonadditive, making it suitable for correlated systems  \cite{Tsallis2003}.

We assume a $q$-Gaussian distribution
\begin{equation}
P_q(x) = \frac{1}{Z_q}\left[1 + (q-1)\beta_q (x - M)^2\right]^{\frac{1}{1-q}},
\tag{20}
\end{equation}
where M is the q-mean \cite{Devi2017}. The normalization is
\begin{equation}
Z_q = \frac{C_q}{\sqrt{\beta_q}},
\tag{20a}
\end{equation}
\begin{equation}
C_q = \sqrt{\pi}\,
\frac{\Gamma\!\left(\frac{1}{q-1} - \frac{1}{2}\right)}
{\sqrt{q-1}\,\Gamma\!\left(\frac{1}{q-1}\right)}.
\tag{20b}
\end{equation}

Using (19)–(20b), the Tsallis entropy becomes
\begin{equation}
S_q = \alpha - (\alpha - 1/2)\left(\frac{\sqrt{\kappa}}{C_\alpha}\right)^{1/\alpha},
\tag{21}
\end{equation}
where
\begin{equation}
\alpha = \frac{1}{q-1},
\tag{21a}
\end{equation}
\begin{equation}
\kappa = \frac{\beta_q}{\alpha},
\tag{21b}
\end{equation}
\begin{equation}
C_\alpha = \sqrt{\pi}\,\frac{\Gamma(\alpha - 1/2)}{\Gamma(\alpha)}.
\tag{21c}
\end{equation}

The EGF equations for $M$, $\alpha$, and $\beta_q$ are
\begin{equation}
\frac{\partial M}{\partial t} = 0,
\tag{22}
\end{equation}
\begin{equation}
\frac{\partial \alpha}{\partial t} = \gamma_q \frac{\partial S_q}{\partial \alpha},
\tag{22a}
\end{equation}
\begin{equation}
\frac{\partial \beta_q}{\partial t} = \gamma_q \frac{\partial S_q}{\partial \beta_q}.
\tag{22b}
\end{equation}

Using (A5) and (A7) from Appendix A, we obtain
\begin{equation}
\frac{\partial \alpha}{\partial t}
= \gamma_q\left[
1 - B\left(1 - \frac{\alpha - 1/2}{2\alpha^2}\right)
+ \left(1 - \frac{1}{2\alpha}\right)B\left(\ln B + \psi(\alpha - 1/2) - \psi(\alpha)\right)
\right],
\tag{23}
\end{equation}
\begin{equation}
\frac{\partial \beta_q}{\partial t}
= -\gamma_q\,\frac{1 - 1/(2\alpha)}{2\beta_q}\,B,
\tag{24}
\end{equation}
where $\psi$ is the digamma function. $B$ as given in (A2) is  
\begin{equation}
B = \left(\frac{\sqrt{\kappa}}{C_\alpha}\right)^{1/\alpha}.
\tag{25}
\end{equation}
Equation (24) may be rewritten as
\begin{equation}
\frac{\partial \beta_q}{\partial t}
= -\gamma_q\,f(q)\,\beta_q^{(q-3)/2},
\tag{26}
\end{equation}
where
\begin{equation}
f(q) = \frac{1}{2}\left(1 - \frac{1}{2\alpha}\right)
\left(\frac{1}{\sqrt{\pi\alpha}\,C_\alpha}\right)^{1/\alpha}.
\tag{26a}
\end{equation}

For $\gamma_q > 0$, $\beta_q$ is a decreasing function of $t$.  
As shown in Appendix A, in the limit $q \to 1$ ($\alpha \to \infty$), 
$\partial S_q/\partial \alpha \to 0$ and $\partial \alpha/\partial t \to 0$, 
and equation (24) reduces to the Gaussian case (14).

\subsubsection*{EGF Equations for Constant $\alpha$ ($q \neq 1$)}

For constant $\alpha$, $\partial \alpha/\partial t = 0$, and $f(q)$ in (26) is independent of time. Therefore (26) can be solved exactly, which gives
\begin{equation}
\beta_q(t)
= \beta_{0q}\left[
1 - \gamma_q\frac{f(q)}{\beta_{0q}^{(5-q)/2}}\,t
\right]^{2/(5-q)},
\tag{27}
\end{equation}
where $\beta_{0q} = \beta_q(t=0)$.

The relaxation time is
\begin{equation}
T_q = \frac{\beta_{0q}^{(5-q)/2}}{\gamma_q f(q)}.
\tag{28}
\end{equation}

For typical values $q \approx 1.5$, $\beta_{0q} \approx 1.5$, $\beta_{0g} \approx 0.5$, and using  $\gamma_{q}$ = $\gamma_{g}$       ,
\begin{equation}
\frac{T_q}{T_g}
= \frac{f(q)\,\beta_{0q}^{(5-q)/2}}{\beta_{0g}}
\approx 33.
\tag{29}
\end{equation}

Figure~1 compares the relaxation time behaviors of Shannon entropy (Gaussian distribution) 
and Tsallis entropy (q-Gaussian distribution) maximizations. 
In producing this graph, we chose initial parameter values corresponding to S\&P 500 stock data 
just before the 2008 crash.  
A larger relaxation time indicates the persistence of information (inverse entropy) over a longer duration.

\section{Numerical Method and Results}

\subsection{Numerical Solutions for the General Case}

The EGF equations (23) and (26) are coupled nonlinear equations that must be solved numerically. 
We use the Euler forward scheme, where the time derivative is replaced by the discrete update
\begin{equation}
\alpha(t_{n+1}) = \alpha(t_n) + h\,\frac{\partial S_q(t_n)}{\partial \alpha},
\tag{30}
\end{equation}
\begin{equation}
\beta_q(t_{n+1}) = \beta_q(t_n) + h\,\frac{\partial S_q(t_n)}{\partial \beta_q}.
\tag{31}
\end{equation}

In writing (30) and (31), we have absorbed $\gamma_q$ into the iteration step $h$.  
The entropy derivatives on the right-hand side of (30) and (31) are given by (A5) and (A7) in Appendix~A.

To study the effect of nonextensivity on relaxation time, we solve (30) and (31) for two sets of initial conditions:

\[
\{q_0 = 1.587,\ \beta_{0q} = 2.865\}, \qquad
\{q_0 = 1.195,\ \beta_{0q} = 0.682\}.
\]

These values correspond to $q$ and $\beta_q$ estimated from one-day stock returns \cite{Devi2017}.  
The first set corresponds to the period just before the 2008 crash (7 November 2008), while the second corresponds to a relatively quiet period (22 December 2003).  
The higher value of $q_0$ in the first set indicates stronger nonextensivity.

The stability of the Euler method depends strongly on the step size $h$.  
It must be sufficiently small so that the solutions smoothly approach their asymptotic values.  
Equations (30) and (31) were iterated over 30,000 steps with $h = 0.001$.  
Stability was checked by halving $h$; the resulting parameter differences were of order $10^{-8}$.

\subsection{Results}

Figure~2 shows the variation of entropy over 30,000 cycles.  
The relaxation time (time to reach maximum entropy) for the more nonextensive period (higher $q_0$) is significantly longer than that for the less nonextensive period.  
This indicates that information (inverse of entropy) persists longer during highly nonextensive periods, allowing predictions to extend over longer horizons.

Figure~3 shows the dynamics of $\beta_q$ over 30,000 cycles.  
Here again, $\beta_q$ for the smaller $q_0$ falls off much more rapidly than for the larger $q_0$.  
It is important to note that the entropy (Figure~2) has not reached its maximum value even after $\beta_q$ reaches zero.  
From equation (21), even when $\beta_q \to 0$, the entropy continues to evolve through $\alpha$.  
Figure~4 shows that $q$ continues to change even after $\beta_q$ reaches zero.  
Thus, the equilibrium state is not reached until several cycles later.  
This effect is especially pronounced during chaotic (highly nonextensive) periods such as just before the 2008 crash.

We now compare the parameters $q$ and $\beta$ estimated from data with those obtained from the model dynamics.  
We first estimate $q$ and $\beta$ from one-day log returns \cite{Devi2017}.  
Figures~5 and 6 show these values over a period beginning just before the 2008 crash (7 November 2008) and extending 1000 days afterward.  
The values decrease until approximately day 650 and then begin to rise again.  
We take the interval between the initial time and the minimum parameter values as the relaxation time of the data.

The initial values of the model parameters $\{q_0 = 1.587,\ \beta_{0q} = 2.865\}$ match those of the first data sample.  
Thus, we can fit the model and data relaxation times to determine the number of cycles per day.  
Figure~2 shows that the model relaxation time is about 26,000 cycles, giving approximately 40 cycles per day.

Figure~7 compares the dynamics of the model parameter $q$ with those estimated from one-day stock returns.  
Up to about 360 days, the model closely follows the data trend, then falls sharply toward equilibrium (maximum entropy).  
The $\beta_q$ dynamics, shown in Figure~8, decreases continuously.  
As discussed in Section~2.3, $\beta_q$ is a decreasing function of $t$.  
The data show an initial increase in $\beta$ during the first few time steps, which is not captured by the model.  
This gives the appearance of a linear shift in the model $\beta_q$.  
Nevertheless, the overall decreasing trend is captured.

\section*{Summary and Conclusions}

We have developed a method to estimate the relaxation time of a nonextensive system such as financial market dynamics using the Euclidean Gradient Flow (EGF) method for maximizing Tsallis entropy.  
The relaxation time is defined as the time required to reach the maximum entropy state.  
Assuming the system follows a $q$-Gaussian distribution and that the dynamics arise from variations in the parameters $q$ and $\beta_q$, explicit EGF equations were derived in the $\{q,\beta_q\}$ space.

A comparison of relaxation times for Shannon and Tsallis entropy maximization shows that the former is very short, consistent with the Efficient Market Hypothesis (EMH), while the latter is much longer.  
This implies that information (related to inverse entropy) persists over longer periods, allowing predictions over longer horizons.

The $\{q,\beta_q\}$ dynamics depend strongly on initial conditions.  
Across several initial values, higher $q$ (greater nonextensivity) corresponds to longer relaxation times.  
A comparison with S\&P 500 data covering the 2008 crash shows that relaxation and prediction times can be as long as 360 days.

An interesting question is how long financial markets take to settle after chaotic periods such as bubbles or crashes.  
The method discussed here may help answer that question.

\section*{Acknowledgements}

Many thanks to Sherman Page for a critical reading of the manuscript.

\appendix

\section*{Appendix A: Euclidean Gradient Flow Equations for the Maximization of Tsallis Entropy}

From equations (21), (21a), (21b), and (21c), the Tsallis entropy for a $q$-Gaussian distribution is
\begin{equation}
S_q = \alpha - (\alpha - 1/2)\,B.
\tag{A1}
\end{equation}

Here
\begin{equation}
B = \left(\frac{\sqrt{\kappa}}{C_\alpha}\right)^{1/\alpha},
\tag{A2}
\end{equation}
\begin{equation}
C_\alpha = \sqrt{\pi}\,\frac{\Gamma(\alpha - 1/2)}{\Gamma(\alpha)}.
\tag{A3}
\end{equation}

Taking the derivative of $S_q$ with respect to $\alpha$,
\begin{equation}
\frac{\partial S_q}{\partial \alpha}
= 1 - B - (\alpha - 1/2)\,\frac{\partial B}{\partial \alpha}.
\tag{A4}
\end{equation}

Using (A1), (A2), and (A3), it is straightforward to show that
\begin{equation}
\frac{\partial S_q}{\partial \alpha}
= 1 - B\left(1 - \frac{\alpha - 1/2}{2\alpha^2}\right)
+ \left(1 - \frac{1}{2\alpha}\right)B\left[\ln(B) + \psi(\alpha - 1/2) - \psi(\alpha)\right].
\tag{A5}
\end{equation}
Here $\psi$ is the digamma function.

Now take the derivative of $S_q$ with respect to $\beta_q$:
\begin{equation}
\frac{\partial S_q}{\partial \beta_q}
= (\alpha - 1/2)\,\frac{\partial B}{\partial \beta_q},
\tag{A6}
\end{equation}
which gives
\begin{equation}
\frac{\partial S_q}{\partial \beta_q}
= -\frac{1 - 1/(2\alpha)}{2\beta_q}\,B.
\tag{A7}
\end{equation}

\subsection*{Limiting Case: $q \to 1$}

We now examine the behavior of $\partial S_q/\partial \alpha$ and $\partial S_q/\partial \beta_q$ in the limit $q \to 1$ ($\alpha \to \infty$).

Using the limiting behaviors

\[
\lim_{\alpha \to \infty} \Gamma(\alpha - 1/2)
\;\to\;
\frac{\Gamma(\alpha)}{\sqrt{\alpha}},
\]

\[
\lim_{\alpha \to \infty} \left[\psi(\alpha) - \psi(\alpha - 1/2)\right]
\;\to\;
\frac{1}{2\alpha},
\]

it follows that $B \to 1$, and therefore
\begin{equation}
\frac{\partial S_q}{\partial \alpha} \to 0,
\tag{A8}
\end{equation}
\begin{equation}
\frac{\partial S_q}{\partial \beta_q} \to -\frac{1}{2\beta_g}.
\tag{A9}
\end{equation}

Hence, the EGF equation in the limit $q \to 1$ becomes
\begin{equation}
\frac{\partial \beta_q}{\partial t}
= \gamma_q\,\frac{\partial S_q}{\partial \beta_q}
\;\to\;
-\frac{\gamma_g}{2\beta_g},
\tag{A10}
\end{equation}
which is identical to equation (14).

\bibliographystyle{unsrt}
\bibliography{references}

\section*{Figures}

\begin{figure}[h!]
\centering
\includegraphics[width=0.8\textwidth]{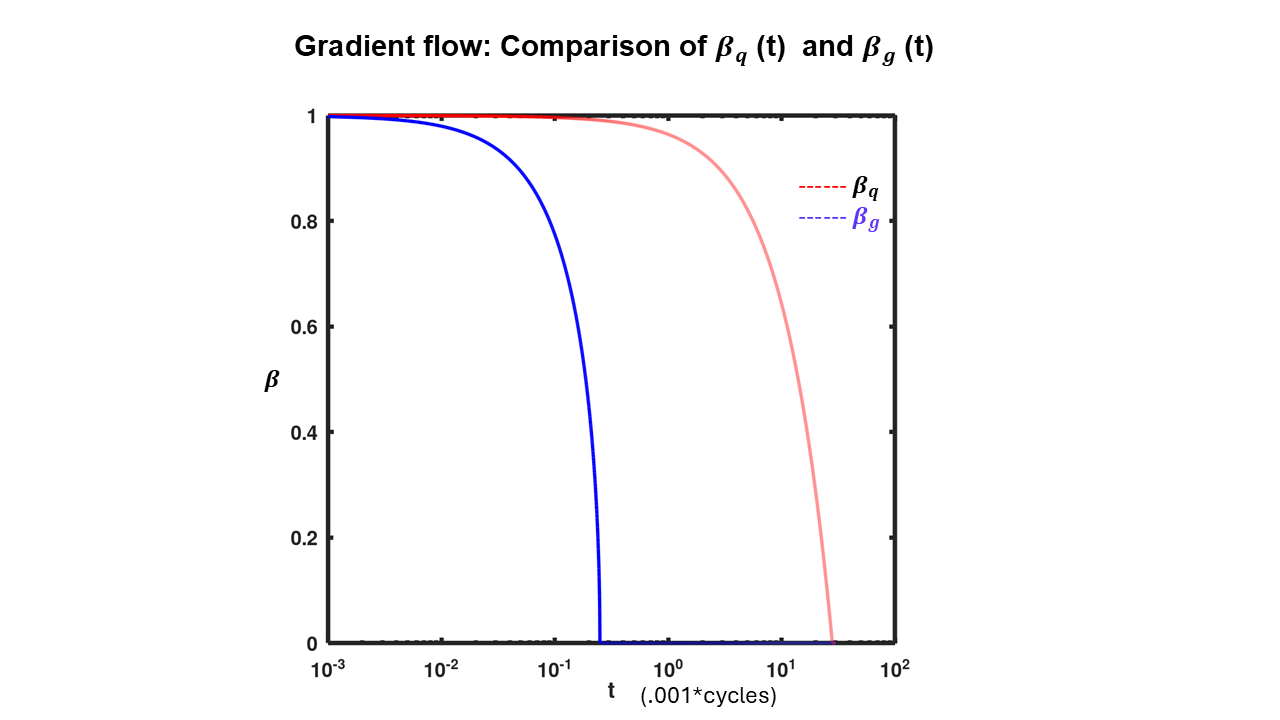}
\caption{Comparison of $\beta_q/\beta_{0q}$ and $\beta_g/\beta_{0g}$ on a logarithmic $t$ axis. The parameters are $q = 1.587$, $\beta_{0q} = 2.864$, and $\beta_{0g} = 0.5$.}
\label{fig:1}
\end{figure}

\begin{figure}[h!]
\centering
\includegraphics[width=0.8\textwidth]{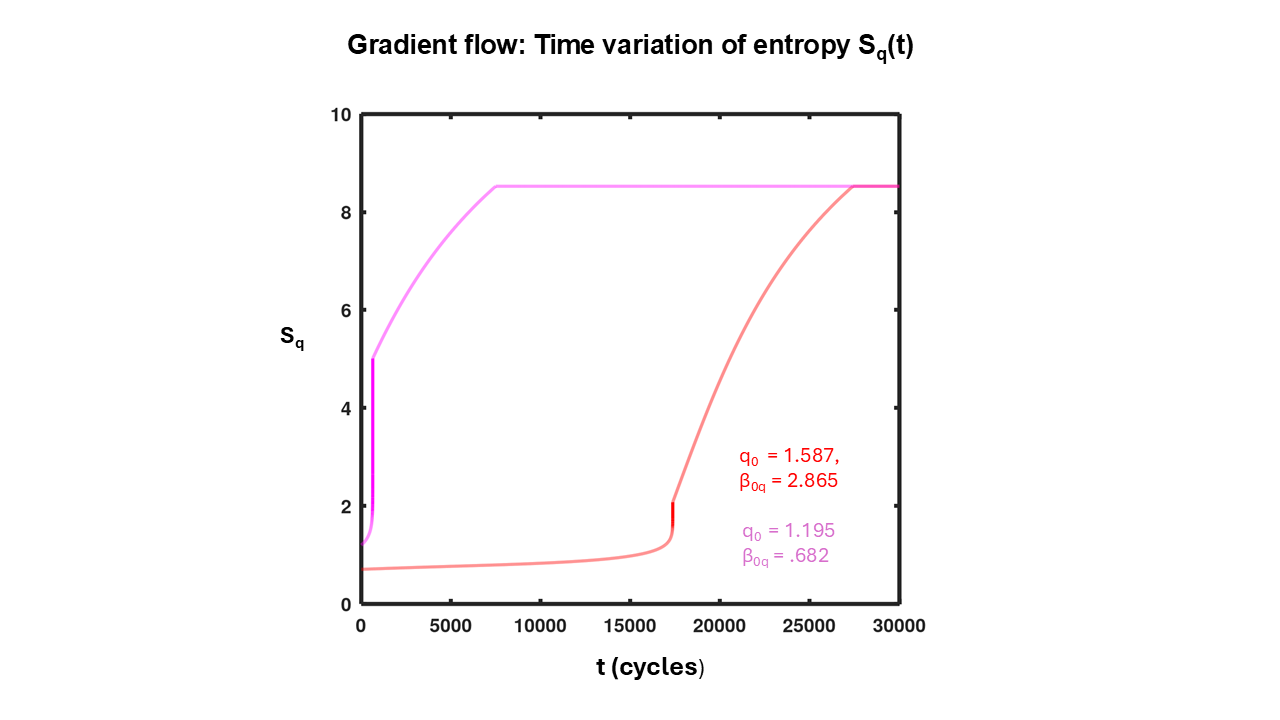}
\caption{Time variation of entropy $S_q(t)$ for two different sets of initial values for $\{q,\beta\}$. The relaxation time is much longer when the nonextensivity is higher (larger $q_0$).}
\label{fig:2}
\end{figure}

\begin{figure}[h!]
\centering
\includegraphics[width=0.8\textwidth]{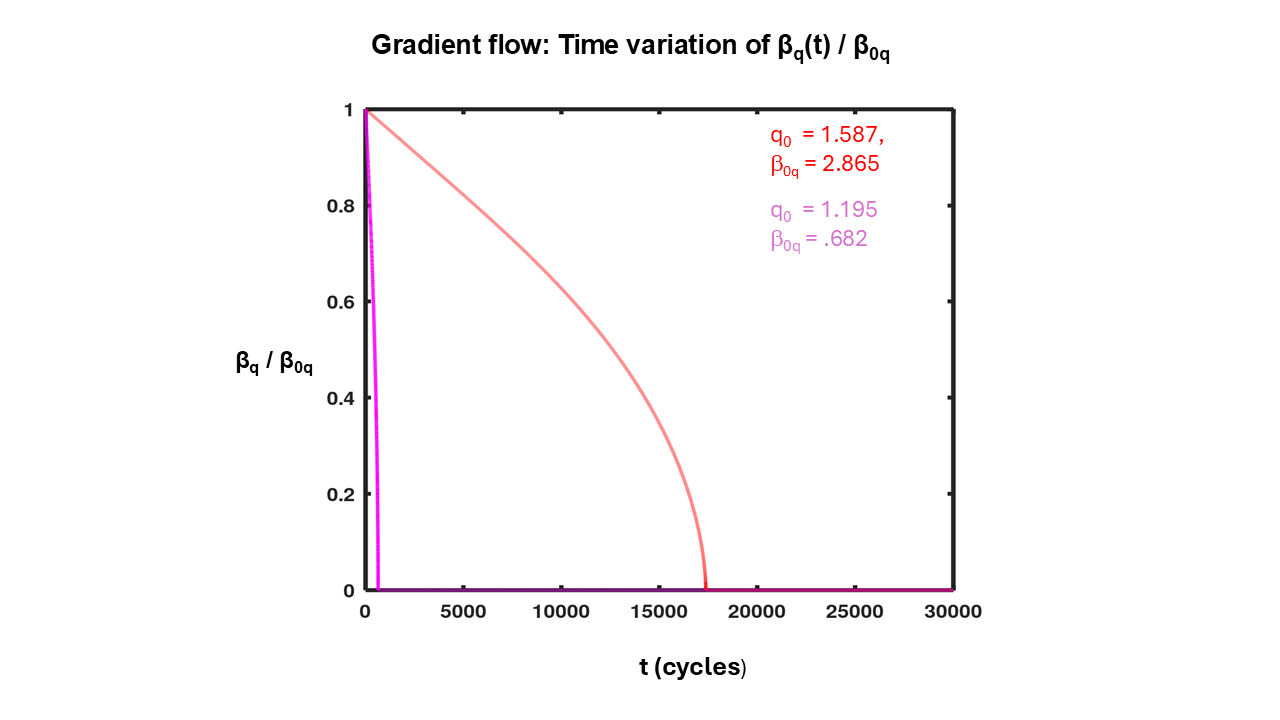}
\caption{Time variation of $\beta_q(t)/\beta_0q$ for two different sets of initial values for $\{q,\beta\}$.}
\label{fig:3}
\end{figure}

\begin{figure}[h!]
\centering
\includegraphics[width=0.8\textwidth]{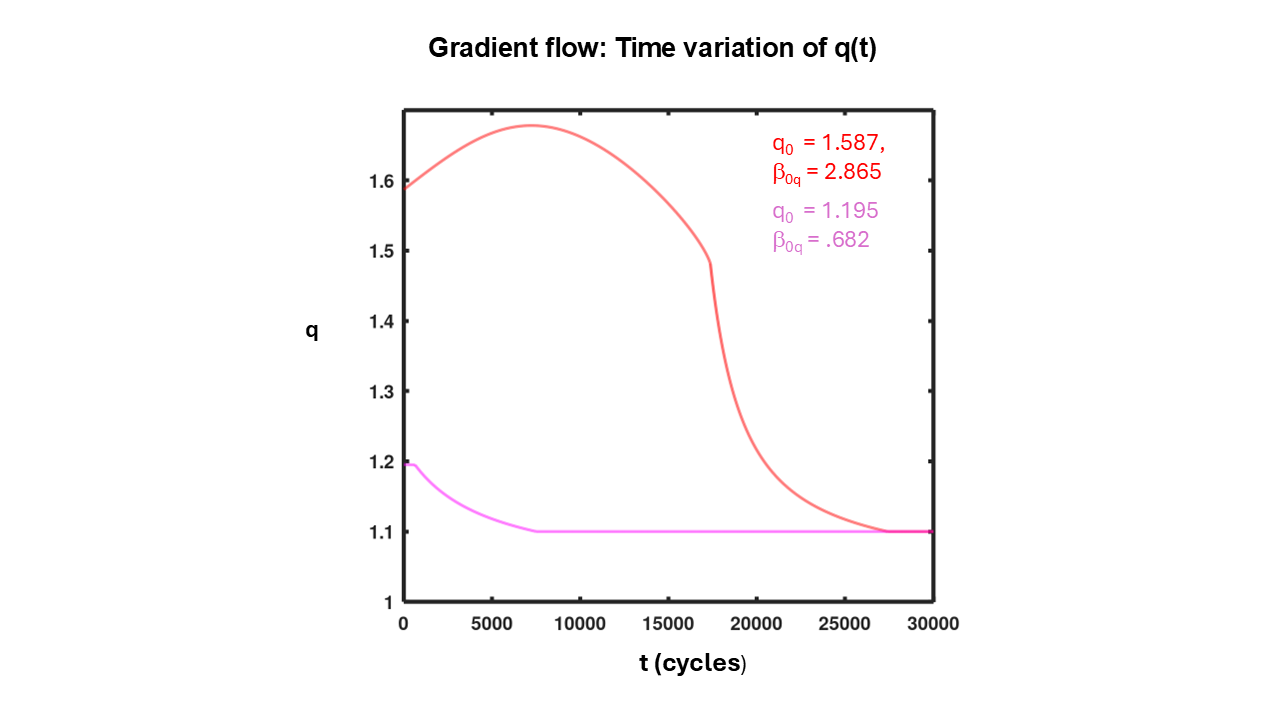}
\caption{Time variation of $q(t)$ for two different sets of initial values for $\{q,\beta\}$.}
\label{fig:4}
\end{figure}

\begin{figure}[h]
\centering
\includegraphics[width=0.80\linewidth]{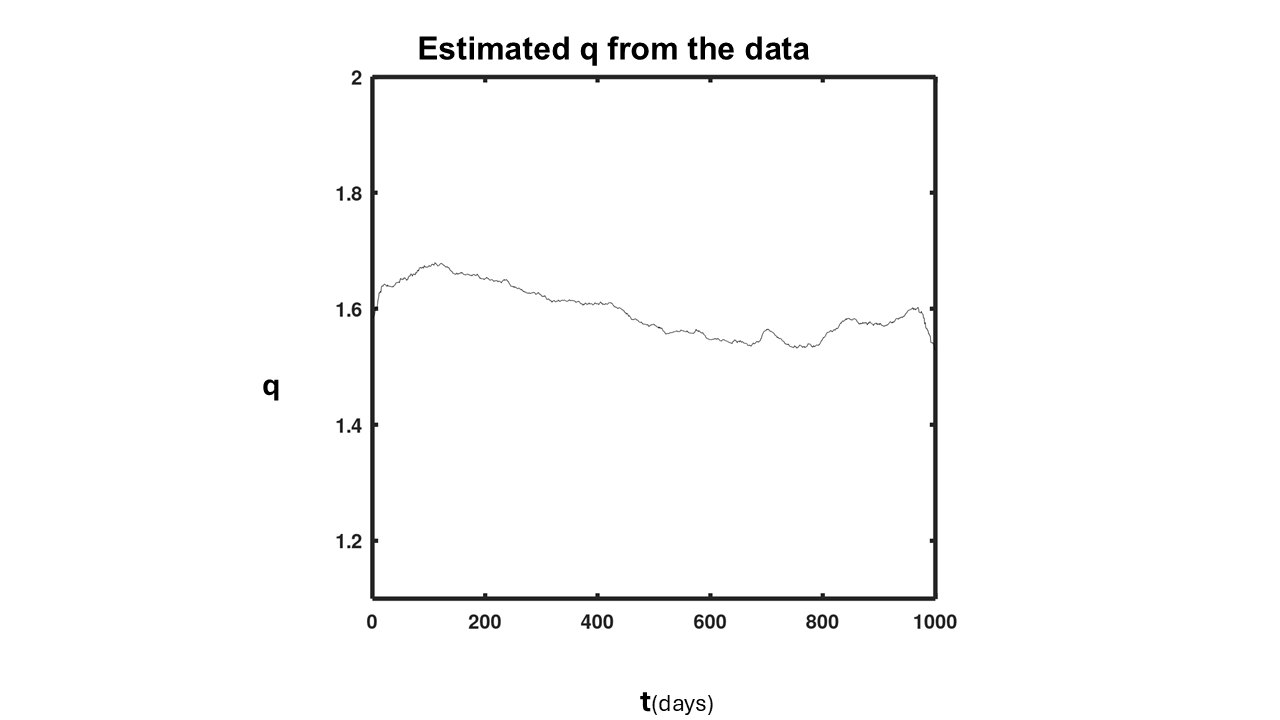}
\caption{Estimated $q$ from one-day log returns. The day range is 1000 samples. 
The first sample corresponds to 7 November 2008.}
\label{fig:q_data}
\end{figure}

\begin{figure}[h]
\centering
\includegraphics[width=0.80\linewidth]{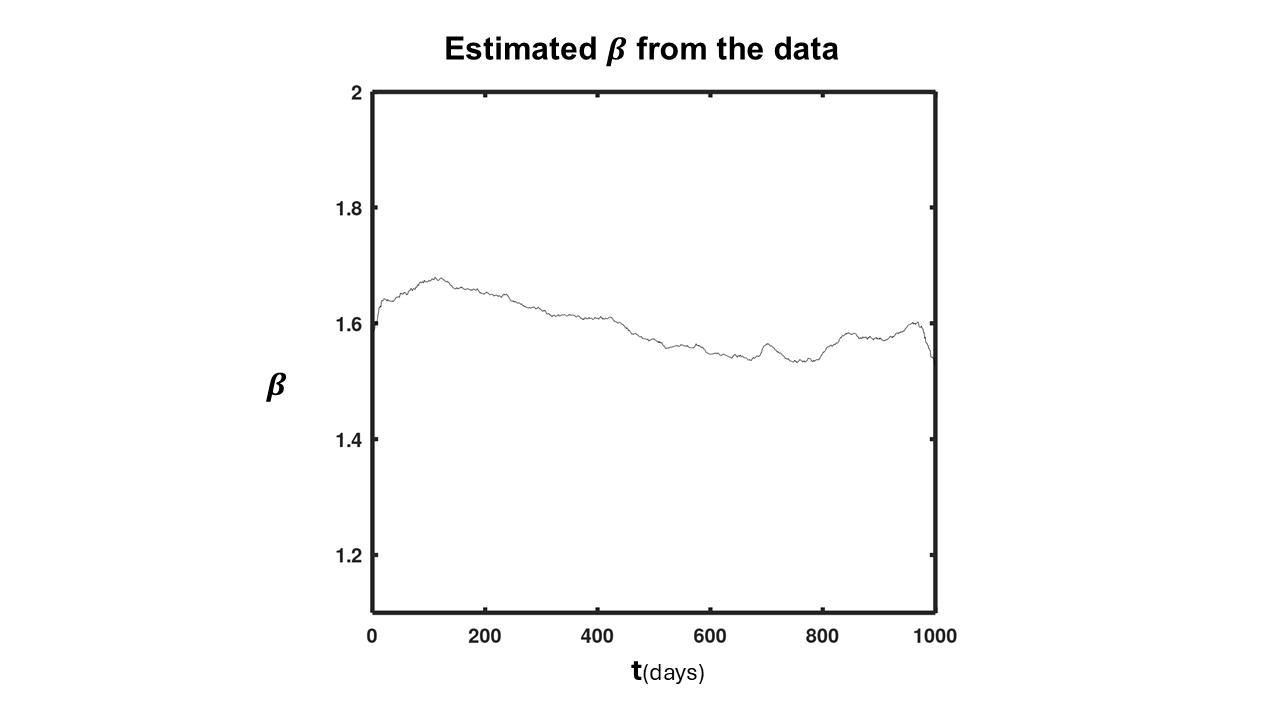}
\caption{Estimated $\beta$ from one-day log returns. The description of the $t$-axis 
is the same as in Figure 5.a}.
\label{fig:beta_data}
\end{figure}

\begin{figure}[h]
\centering
\includegraphics[width=0.80\linewidth]{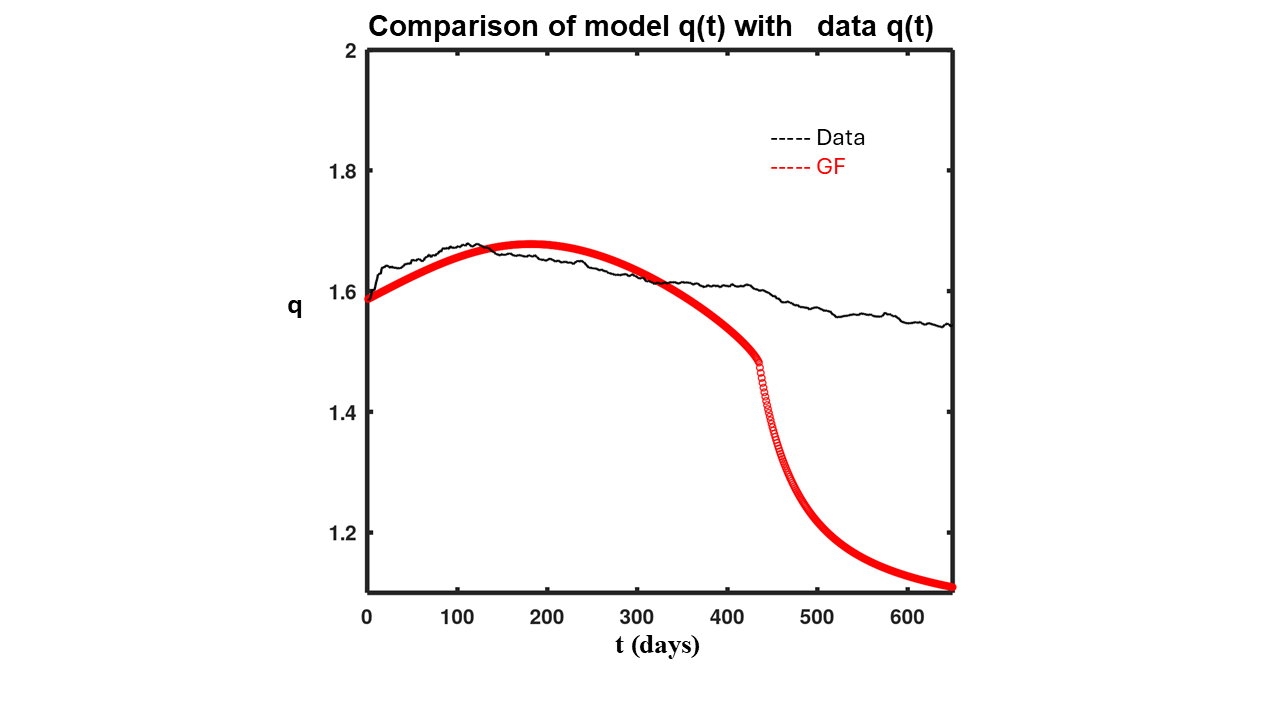}
\caption{Comparison of model $q(t)$ with data $q(t)$ (from one-day returns). 
The day range is 650 samples. The first sample corresponds to 7 November 2008.}
\label{fig:q_compare}
\end{figure}

\begin{figure}[h]
\centering
\includegraphics[width=0.80\linewidth]{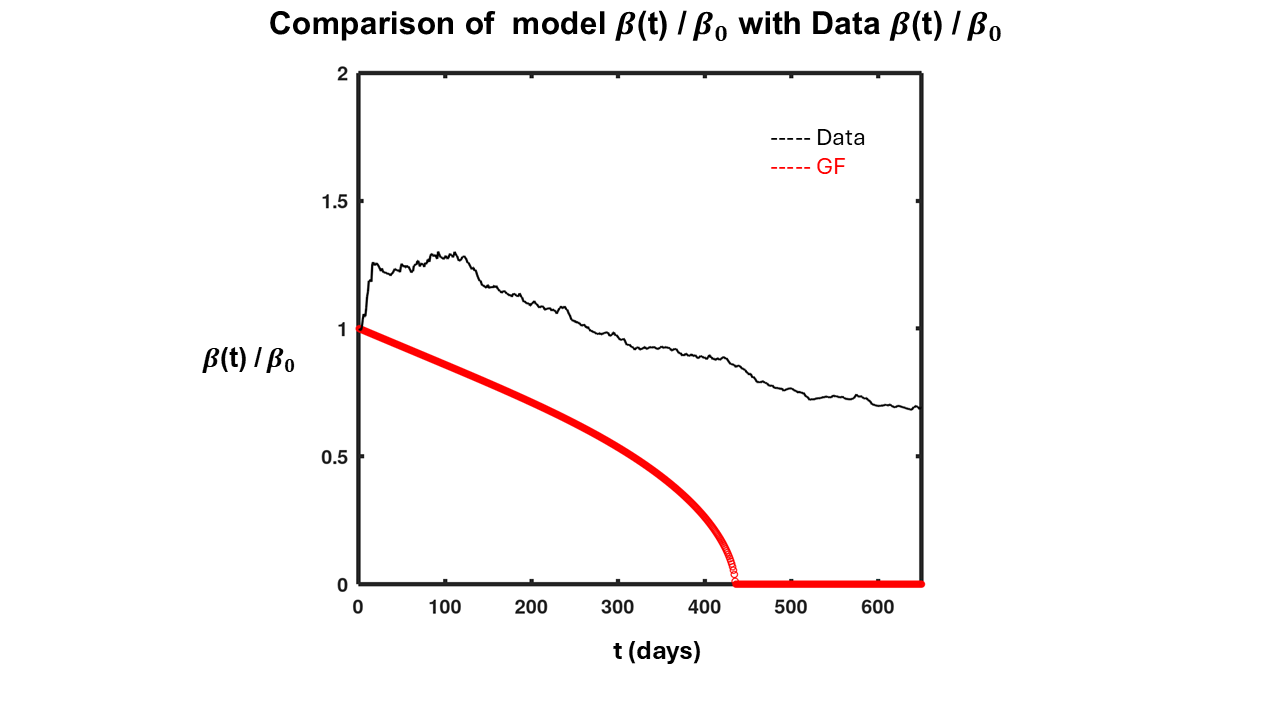}
\caption{Comparison of model $\beta/\beta_{0}$ (same as $\beta_q/\beta_{0q}$)  and data $\beta/\beta_0$. The description of the $t$-axis is the same as in Figure 7.}
\label{fig:beta_compare}
\end{figure}

\end{document}